\documentclass[pra,twocolumn,floatfix,superscriptaddress,longbibliography]{revtex4-1}
\usepackage{tikz}
\usepackage[noend]{algpseudocode}
\usepackage[ruled]{algorithm}
\usepackage{mathtools}
\usetikzlibrary{shapes,arrows,positioning}
\usepackage{standalone}
\usepackage{amssymb,amsmath,amstext}
\usepackage{graphicx}
\usepackage{epstopdf}
\usepackage{color}
\usepackage{appendix}
\usepackage[T1]{fontenc}
\usepackage{bbold}
\usepackage{bbm}
\usepackage{float}
\usepackage{latexsym}
\usepackage{xr-hyper}
\usepackage[colorlinks=true,citecolor=blue,linkcolor=magenta]{hyperref}
\usepackage[latin1]{inputenc}

\usepackage{enumitem}

\usepackage{hhline}
 
\usetikzlibrary{shapes,arrows}

\renewcommand{\vec}[1]{\boldsymbol{#1}}  

\long\def\ca#1\cb{} 

\newcommand{\braket}[2]{\langle #1 \hspace{1pt} | \hspace{1pt} #2 \rangle}

\newcommand{\ketbra}[2]{| \hspace{1pt} #1 \rangle \langle #2 \hspace{1pt} |}

\newcommand{\ket}[1]{|#1\rangle}               
\newcommand{\bra}[1]{\langle #1|}              
\newcommand{\dya}[1]{\ket{#1}\!\bra{#1}}
\newcommand{\matl}[3]{\langle #1|#2|#3\rangle} 

\def\NoNumber#1{{\def\alglinenumber##1{}\State #1}\addtocounter{ALG@line}{-1}}

\newcommand{\rank}{\text{rank}}

\newcommand{\LC}{\mathcal{L}}

\newcommand{\OC}{\mathcal{O}}

\newcommand{\SC}{\mathcal{S}}

\newcommand{\ZC}{\mathcal{Z}}

\newcommand{\Tr}{{\rm Tr}}

\newcommand{\ave}[1]{\langle #1\rangle}               
\renewcommand{\geq}{\geqslant}
\renewcommand{\leq}{\leqslant}

\renewcommand{\vec}[1]{\boldsymbol{#1}}  

\newcommand{\ad}{^\dagger}

\newcommand*{\id}{\openone}

\newcommand{\rhot}{\tilde{\rho}}

\renewcommand{\th}{\theta } 

\newcommand{\thv}{\vec{\theta}}
\newcommand{\zv}{\vec{z}}
\newcommand{\ev}{\vec{e}}
\newcommand{\pv}{\vec{p}}
\newcommand{\opt}{\text{opt}}
\newcommand{\poly}{\text{poly}}
\newcommand{\thvo}{\vec{\theta}_{\opt}}


   \makeatletter
     \renewcommand\@make@capt@title[2]{%
      \@ifx@empty\float@link{\@firstofone}{\expandafter\href\expandafter{\float@link}}%
       {\textbf{#1}}\@caption@fignum@sep#2\quad}%
     \makeatother
 
\makeatletter 
\renewcommand{\fnum@figure}{\textbf{Figure~\thefigure}}
\makeatother

\begin{document}

\title{Variational Quantum State Eigensolver}

\author{M. Cerezo}
\thanks{Corresponding author: \texttt{cerezo@lanl.gov}}
\affiliation{Theoretical Division, MS B213, Los Alamos National Laboratory, Los Alamos, NM 87545, USA.}
\affiliation{Center for Nonlinear Studies, Los Alamos National Laboratory, Los Alamos, NM, USA
}

\author{Kunal Sharma}
\affiliation{Theoretical Division, MS B213, Los Alamos National Laboratory, Los Alamos, NM 87545, USA.}
\address{Hearne Institute for Theoretical Physics and Department of Physics and Astronomy, Louisiana State University, Baton Rouge, LA USA.}

\author{Andrew Arrasmith}
\affiliation{Theoretical Division, MS B213, Los Alamos National Laboratory, Los Alamos, NM 87545, USA.}

\author{Patrick J. Coles} 
\affiliation{Theoretical Division, MS B213, Los Alamos National Laboratory, Los Alamos, NM 87545, USA.}

\begin{abstract}
Extracting eigenvalues and eigenvectors of exponentially large matrices will be an important application of near-term quantum computers. The Variational Quantum Eigensolver (VQE) treats the case when the matrix is a Hamiltonian. Here, we address the case when the matrix is a density matrix $\rho$. We introduce the Variational Quantum State Eigensolver (VQSE), which is analogous to VQE in that it variationally learns the largest eigenvalues of $\rho$ as well as a gate sequence $V$ that prepares the corresponding eigenvectors. VQSE exploits the connection between diagonalization and majorization to define a cost function $C=\Tr(\tilde{\rho} H)$ where $H$ is a non-degenerate Hamiltonian. Due to Schur-concavity, $C$ is minimized when $\tilde{\rho} = V\rho V^\dagger$ is diagonal in the eigenbasis of $H$. VQSE only requires a single copy of $\rho$ (only $n$ qubits) per iteration of the VQSE algorithm, making it amenable for near-term implementation. We heuristically demonstrate two applications of VQSE: (1) Principal component analysis, and (2) Error mitigation. 
\end{abstract}

\maketitle
\onecolumngrid
\vspace{-.4cm}
\vspace{.4cm}

\twocolumngrid

\section{Introduction}

Near-term quantum computers hold great promise but also pose great challenges. Low qubit counts place constraints on problem sizes that can be implemented. Decoherence and gate infidelity place constraints on the circuit depth that can be implemented. These constraints are captured in the (now widely used) term Noisy Intermediate-Scale Quantum (NISQ)~\cite{preskill2018quantum}.

To address the circuit depth constraint, Variational Quantum Algorithms (VQAs) have been proposed for many applications~\cite{peruzzo2014VQE, farhi2014QAOA, johnson2017qvector, romero2017quantum, larose2018, arrasmith2019variational, cerezo2019variational, jones2019variational, yuan2018theory, li2017efficient, kokail2019self, Khatri2019quantumassisted, jones2018quantum, heya2018variational, endo2018variational,sharma2019noise, carolan2019variational,yoshioka2019variational,bravo-prieto2019,xu2019variational,mcardle2019variational,cirstoiu2019variational,otten2019noise,LubaschVariational20,verdon2019quantum,bravo2019quantum}. VQAs employ a quantum-classical optimization loop to train the parameters $\thv$ of a quantum circuit $V(\thv)$. Leveraging classical optimizers allows the quantum circuit depth to remain shallow. This makes VQAs powerful tools for error mitigation on NISQ devices.

A particularly important application of NISQ computers will be extracting the spectra, eigenvalues and eigenvectors, of very large matrices. Indeed the most famous VQA, known as the Variational Quantum Eigensolver (VQE), aims to variationally determining the energies and state-preparation circuits for the ground state and low-lying excited states of a given Hamiltonian, i.e., a Hermitian matrix. VQE promises to revolutionize the field of quantum chemistry~\cite{cao2019quantum,mcardle2018quantum}, and perhaps even nuclear~\cite{dumitrescu2018cloud} and condensed matter~\cite{wecker2015progress,bauer2016hybrid} physics.

If one instead considers a positive-semidefinite matrix, then extracting the spectrum has direct application as a machine-learning primitive known as Principal Component Analysis (PCA). Along these lines, Lloyd et al.~\cite{lloyd2014quantum} introduced a quantum algorithm called quantum PCA (qPCA) to deterministically extract the spectrum of an $n$-qubit density matrix $\rho$. qPCA employs quantum phase estimation and density matrix exponentiation as subroutines and hence requires a large number of quantum gates and copies of $\rho$. In an effort to reduce circuit depth in the NISQ era, LaRose {\it et al}.~\cite{larose2018} developed a VQA for this application called Variational Quantum State Diagonalization (VQSD). VQSD requires two copies of $\rho$, hence $2n$ qubits, and  trains the parameters $\thv$ of a gate sequence $V(\thv)$ so that $\rhot = V(\thv)\rho V\ad(\thv)$ is approximately diagonal. A different variational approach, called Quantum Singular Value Decomposition (QSVD), was introduced by Bravo-Prieto {\it et al}.~\cite{bravo2019quantum}. QSVD takes a purification $\ket{\psi}$ of $\rho$ as its input and hence requires however many qubits it takes to purify $\rho$ (possibly $2n$ qubits).

In this work, we introduce a variational algorithm for PCA that only requires a single copy of $\rho$ and hence only $n$ qubits per iteration of the algorithm.  Our approach, called the Variational Quantum State Eigensolver (VQSE), exploits the mathematical connection between diagonalization and majorization. Namely, it is well known that the eigenvalues of a density matrix $\rho$ majorize the diagonal elements in any basis. Hence, by choosing a cost function $C$ that is a Schur concave function of the diagonal elements of $\rho$, one can ensure that the cost function is minimized when $\rho$ is diagonalized. Specifically, we write the cost as $C = \Tr(\rhot H)$, where $H$ is some Hamiltonian with a non-degenerate spectrum, which ensures the Schur concavity property. Note that evaluating $C$ simply involves measuring the expectation value of $H$ on $\rhot$, and hence one can see why only $n$ qubits are required.

To learn the optimal $\thv$ parameters, we introduce a new training approach, not previously used in other VQAs. Specifically, we employ a time-dependent Hamiltonian $H$ that we adapt based on information gained from measurements performed throughout the optimization. The aim of this adaptive approach is: (1) to mitigate barren plateaus in training landscapes, and (2) to get out of local minima. With our numerics, we find that using an adaptive Hamiltonian is better than simply fixing the Hamiltonian throughout the optimization. Here, we further provide a rigorous analysis of the measurement shot requirements of VQSE where we show that the relative error induces from statistical sampling error is, with high probability, smaller than $\delta$, if one measures the system response with a number of shots that scales only as $\Omega(\log(1/\delta)/\lambda_m^2)$, with $\lambda_m$ being the smallest eigenvalue one wishes to estimate.

Finally, we illustrate two important applications of VQSE with our numerical implementations. First,  we use VQSE for error mitigation of the $W$-state preparation circuit. Namely, by projecting the state onto the eigenvector with the largest eigenvalue, we re-purify the state, mitigating the effects of incoherent errors.  Second,  we use VQSE to 
perform entanglement spectroscopy (which is essentially PCA on the reduced state of a bipartition) on the ground state of an $XY$-model spin chain. This allows us to identify quantum critical points in this model.

\section{Results}

\subsection{Theoretical Basis of VQSE}\label{subsec:theo}

Consider an $n$-qubit quantum state $\rho$ with (unknown) spectral decomposition
$\rho=\sum_k \lambda_k\dya{\lambda_k}$, such that the eigenvalues are ordered in decreasing order (i.e., $\lambda_k\geq\lambda_{k+1}$ for $k=1,\ldots,\rank(\rho)$, while $\lambda_k=0$ for $k\geq \rank(\rho)$).  The goal of VQSE is to estimate the  $m$-largest eigenvalues of $\rho$, where $m \ll 2^n$, and furthermore to return a gate sequence $V(\thv)$ that approximately prepares their associated eigenvectors from standard basis elements.

At first sight, this looks like a matrix diagonalization problem. Indeed, this is the perspective taken in the literature, e.g., by the VQSD algorithm~\cite{larose2018} which employs a cost function that quantifies how far $\rhot = V(\thv)\rho V\ad(\thv)$ is from a diagonal matrix. However, our VQSE algorithm takes a conceptually different approach, focusing on majorization instead of diagonalization.

We write the VQSE cost function as an energy, or the expectation value of a Hamiltonian:
\begin{equation}\label{eq:cost}
    C(\thv) \equiv \ave{H} = \Tr\left[H V(\thv)\rho V\ad(\thv)\right]\,.
\end{equation}
Here,  $H$ is a simple $n$-qubit Hamiltonian that is diagonal in the standard basis and whose eigenenergies and associated eigenstates are known and respectively given by $\{E_k\}$ and $\{\ket{\ev_k}\}$ (where $\ev_k=e_k^1\cdot\ldots\cdot e_k^n$ for $k=1,\ldots,2^n$ are bitstrings of length $n$). Moreover, we henceforth assume that  the eigenenergies are non-negative and ordered in increasing order, i.e., $E_k\leq E_{k+1}$. We have
\begin{equation}
C(\thv)=\sum_{k=1}^{2^n}E_k p_k=\vec{E}\cdot\pv\,, \quad p_k= \matl{\ev_k}{\rhot}{\ev_k}\,,
\end{equation}
where we  defined the vectors $\vec{E}=(E_1,E_2,\ldots)$ and
$\pv=(p_1,p_2,\ldots)$.
Similarly, let us define the vector of eigenvalues of $\rho$ as $\vec{\lambda}=(\lambda_1,\lambda_2,\ldots)$. Then, since the eigenvalues of a positive semidefinite matrix majorize its diagonal elements $\vec{\lambda}\succ \vec{p}$, and since the dot product with an increasingly ordered vector is a Schur concave function~\cite{horn1990matrix,bhatia2013matrix}, we have
\begin{equation}\label{eq:mincost}
    C(\thv)=\vec{E}\cdot\pv\geq \vec{E}\cdot\vec{\lambda}=\sum_{k}E_k \lambda_{k}\,,
\end{equation}
where we have used the fact that $\rho$ and $\rhot$ have the same eigenvalues. Hence, one can see that $C(\thv)$ is minimized when $V(\thv)$ maps the eigenbasis of $\rho$ to the eigenbasis of $H$, with appropriate ordering. Since the latter is chosen to be the standard basis, this corresponds to diagonalizing $\rho$. Thus, even though it may not be obvious at first sight, minimizing $C(\thv)$ corresponds to diagonalizing $\rho$.

\begin{figure}[t]
    \centering
    \includegraphics[width=\columnwidth]{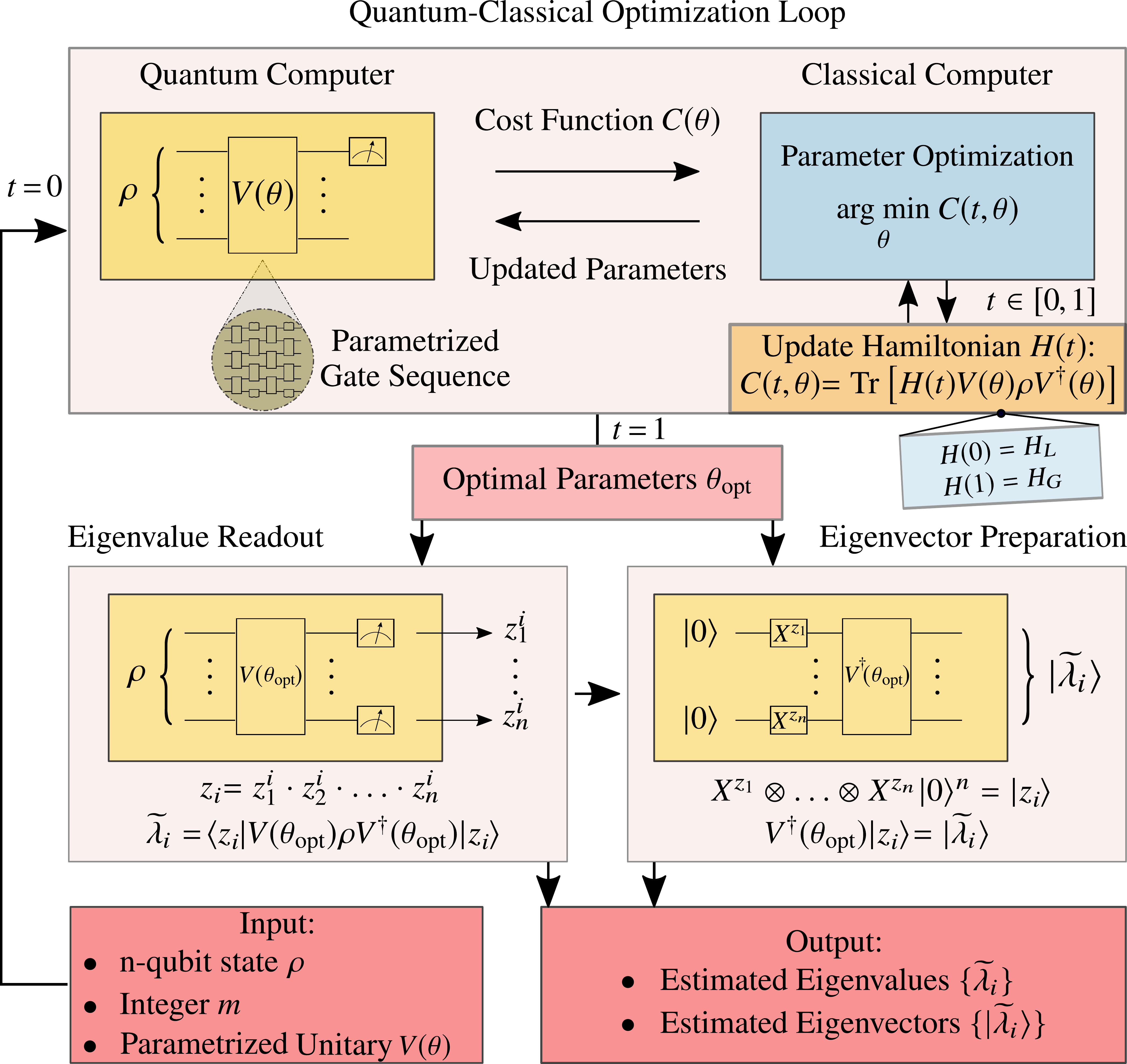}
    \caption{\textbf{Schematic diagram of VQSE}. VQSE takes as inputs an $n$-qubit state $\rho$,  an integer $m$, and a parametrized unitary $V(\thv)$. It then outputs estimates of the  $m$-largest eigenvalues of $\rho$, and their associated eigenvectors. The first step of the algorithm is a hybrid quantum-classical optimization loop to train the parameters $\thv$, and minimize the cost function defined in \eqref{eq:costth} as the expectation value of a Hamiltonian $H(t)$ over the state $\rhot = V(\thv)\rho V\ad(\thv)$. To facilitate this optimization, we adaptively update $H(t)$ using information obtained via measurements on $\rhot$.  When this optimization terminates, at which point we say $\thv = \thvo$, one reads off the eigenvalues. Namely, by preparing $V(\thvo)\rho V\ad(\thvo)$ and measuring in the standard basis, one obtains bitstrings $\zv$ whose associated frequencies are estimates of the eigenvalues of $\rho$. Finally, one prepares the estimated eigenvectors by preparing the states $\ket{\zv}$ and acting on them with $V\ad(\thvo)$.    }   \label{fig:VQSE}
\end{figure}

\subsection{The VQSE algorithm}

Figure~\ref{fig:VQSE} shows a schematic diagram of the Variational Quantum State Eigensolver (VQSE) algorithm. The three inputs to  VQSE are: (1) a $n$-qubit quantum state $\rho$, (2) an integer $m$, and (3) a parameterized gate sequence or ansatz $V(\thv)$. The outputs of VQSE are: (1) estimates $\{\tilde{\lambda}_i\}_{i=1}^m$ of the $m$-largest eigenvalues $\{\lambda_i\}_{i=1}^m$ of $\rho$, and (2) a gate sequence $V(\thv_{\opt})$ that prepares approximate versions $\{\ket{\tilde{\lambda}_i}\}_{i=1}^m$ of the associated $m$ eigenvectors $\{\ket{\lambda_i}\}_{i=1}^m$. While in principle $m$ can be as large as $2^n$, we assume that one is interested in a number $m$ of eigenvalues and eigenvectors that grows at worse as $\OC(\poly(n))$.

After taking in the inputs, VQSE enters a hybrid quantum-classical optimization loop to train the parameters $\thv$ in the ansatz $V(\thv)$. This loop employs a quantum computer to evaluate the VQSE cost function, denoted 
\begin{equation}\label{eq:costth}
    C(t,\thv) \equiv \ave{H(t)} = \Tr\left[H(t)\widetilde{\rho}\right]\,, \quad \widetilde{\rho}=V(\thv)\rho V\ad(\thv)\,.
\end{equation}
Here, $H(t)$ is a Hamiltonian that could, in general, depend on the time $t$, where $t\in[0,1]$ is a parameter that indicates the optimization loop run-time such that the loop starts at $t=0$ and ends at $t=1$. For all $t$, we assume that $H(t)$ can be efficiently measured on a quantum computer and that it is diagonal in the standard basis, with its lowest $m$ eigenenergies being non-degenerate and non-negative. We further elaborate on how to choose $H(t)$ in Section~\ref{sec:cost-funtions}. Note that the quantum circuit to evaluate the cost $C(t,\thv)$, as depicted in Fig.~\ref{fig:VQSE}, simply involves applying $V(\thv)$ to the state $\rho$ and then measuring the Hamiltonian $H(t)$.

The quantum computer then feeds the value of the cost (or the gradient of the cost for gradient-based optimization) to a classical computer, which adjusts the parameters $\thv$ for the next round of the loop. The ultimate goal is to find the global minimum of the cost landscape at $t=1$, i.e., to solve the problem:
\begin{equation}\label{eq:thetopt}
\thv_{\opt}\equiv \arg \min_{\thv} C (1,\thv)\,.
\end{equation}
In reality, one will need to impose some termination condition on the optimization loop and hence the final parameters obtained (which we still denote as $\thv_{\opt}$) will only approximately satisfy Eq.~\eqref{eq:thetopt}. Nevertheless, we provide a verification procedure below in Section~\ref{sec:verif} that allows one to quantify the quality of the solution even when \eqref{eq:thetopt} is not exactly satisfied.

As shown in Fig.~\ref{fig:VQSE}, the next step of  VQSE is the eigenvalue readout. From the parameters $\thvo$ one can estimate the eigenvalues of $\rho$ by acting with the gate sequence $V(\thvo)$ and then measuring in the standard basis $\{\ket{\zv_k}\}$. Let $\Pr(\zv_k)$ be the probability of the $\zv_k$ outcome. Then by taking the $m$ largest of these probabilities we define $\LC \equiv \{\widetilde{ \lambda}_i\}_{i=1}^m$ as the ordered set of estimates of the $m$-largest eigenvalues of $\rho$, and we define $\ZC$ as the set of bitstrings $\{\zv_i\}_{i=1}^m$ associated with the elements of $\LC$:
\begin{equation}\label{eq:lambdaest}
   \widetilde{ \lambda}_i=\Pr(\zv_i)=\matl{\zv_i}{\widetilde{\rho}}{\zv_i}\,,\quad \text{such that} \quad \widetilde{ \lambda}_i\geq \widetilde{ \lambda}_{i+1}\,.
\end{equation}
Note that $\widetilde{ \lambda}_i$   in~\eqref{eq:lambdaest} correspond to diagonal elements of $\widetilde{\rho}$ in the standard basis, and not to its eigenvalues.

In practice, when estimating the eigenvalues one measures $\rhot$ in the standard basis a finite number of times $N_{\text{runs}}$. Hence, if a  bitstring $\zv_i\in\ZC$ has frequency $f_i$ for $N_{\text{runs}}$ total runs, then  we can estimate $\widetilde{ \lambda}_i$ as
\begin{equation}\label{eq:esteigenv}
    \widetilde{ \lambda}_i^{\text{est}}=\frac{f_i}{N_{\text{runs}}}\,.
\end{equation}
One can think of this as a Bernouilli trial. Let $\Lambda_i$ be a random variable that takes value $1$ if we get outcome $\zv_i$ (with probability $\widetilde{ \lambda}_i$), and takes value $0$ otherwise (with probability $1-\widetilde{ \lambda}_i$). After repeating the experiment $N_{\text{runs}}$ times we are interested in bounding the probability that the relative error $\varepsilon_i\equiv |\widetilde{ \lambda}_i^{\text{est}}-\widetilde{ \lambda}_i|/\widetilde{ \lambda}_i$ is larger than a certain value $c\geq 0$. From Hoeffding's inequality, we find
\begin{equation}\label{eq:boundprob}
    \Pr(\varepsilon_i\geq c)\leq e^{-2N_{\text{runs}}c^2\widetilde{\lambda}_i^2}\,, \quad  \forall c>0\,.
\end{equation}
For fixed $N_{\text{runs}}$, Eq.~\eqref{eq:boundprob} shows that the smaller the inferred eigenvalue $\widetilde{ \lambda}_i$, the larger the probability of having a given relative error. Equation~\eqref{eq:boundprob} also implies that increasing $N_{\text{runs}}$ reduces the probability of large relative errors. Hence, we can always choose $N_{\text{runs}}$ such that the probability of error is smaller than a given $\delta$  for all $m$ eigenvalues via
\begin{equation}\label{eq:runsestimates}
    \forall i\in [1,m],\,\,\,\Pr(\varepsilon_i\geq c)\leq \delta \, \rightarrow \, N_{\text{runs}}\geq \frac{\log(1/\delta)}{2c^2\lambda_m^2}\,,
\end{equation}
where $\lambda_m$ is the smallest eigenvalue of interest. Analogously, from~\eqref{eq:runsestimates} we have that all eigenvalues larger than $\sqrt{ \frac{\log(1/\delta)}{2c^2N_{\text{runs}}}}$ have a probability of error smaller than $\delta$.

The last step of VQSE is to prepare the inferred eigenvectors of $\rho$. Given a bitstring $\zv_i\in\ZC$, one can prepare the associated inferred eigenvector by taking the state $\ket{\vec{0}}=\ket{0}^{\otimes n}$, acting on it with the gate $X^{z_1^i}\otimes X^{z_2^i}\otimes\ldots\otimes X^{z_n^i}$, and then applying the gate sequence $V(\thvo)\ad$:
\begin{equation}
     \ket{\widetilde{ \lambda}_i}=V\ad(\thvo)\ket{\zv_i}\,, \quad \ket{\zv_i}=X^{z_1^i}\otimes\ldots\otimes X^{z_n^i}\ket{\vec{0}}\,.
\end{equation}
Note that while the inferred eigenvalues can be stored classically, the eigenvectors are prepared on a quantum computer, and hence one needs to perform measurements to extract information about these eigenvectors.

\subsection{Cost functions }\label{sec:cost-funtions}

Consider the Hamiltonian $H(t)$ that defines the VQSE cost function in \eqref{eq:costth}. Recall that we choose $H(t)$ so that: (1) it is diagonal in the standard basis,  (2) its lowest $m$ eigenvalues are non-negative and non-degenerate, and (3) it can be efficiently measured on a quantum computer. Let us now discuss possible choices for $H(t)$.

\textbf{Fixed Hamiltonians}. When the Hamiltonian is fixed (i.e., time-independent), we write  $H(t)\equiv H$, and $C(t,\thv)\equiv C(\thv)$. In this case, a simple, intuitive cost function is given by
\begin{equation}\label{eq:costglobalm}
    C_G(\thv)=\Tr[H_G\widetilde{\rho}]\,, \quad H_G=\id-\sum_{i=1}^{m} q_i \dya{\vec{e}_i}\,, 
\end{equation}
with $q_i>0$ (such that $q_i> q_{i+1}$),  and where the  $\ket{\vec{e}_i}$ are orthogonal states in the standard basis. The spectrum of $H_G$ is composed of $m$ non-degenerate eigenenergies, and a $(2^n-m)$-fold degenerate eigenenergy.

On the one hand, this large degeneracy makes it easier to find a global minimum as the solution space is large. That is, denoting as $V_{\opt}$  an optimal unitary that minimizes~\eqref{eq:costglobalm}, then there is a large set of such optimal unitaries $\SC_{\text{\opt}}=\{V_{\text{\opt}}\}$, which are not related by global phases. This is due to the fact that one is only interested in the $m$ rows and the $m$ columns of $V(\thv)$ that diagonalize $\widetilde{\rho}$ in the subspace spanned by $\{\ket{\ev_i}\}_{i=1}^m$. Specifically, any optimal unitary must satisfy $\matl{\zv_i}{V_{\text{opt}}}{\lambda_i}=\matl{\lambda_i}{V_{\text{opt}}}{\zv_i}=\delta_{\zv_i\ev_i}$ for $i=1,\ldots, m$  (and with $\zv_i\in\ZC$), while the $(2^n-m)\times(2^n-m)$ unitary principal submatrix of $V_{\text{opt}}$ with matrix elements $\matl{\zv_i}{V_{\text{opt}}}{\zv_{i'}}$, where $\zv_i,\zv_{i'}\not\in\ZC$, remains completely arbitrary.

On the other hand, it has been shown that when employing hardware-efficient ansatzes~\cite{kandala2017} for $V(\thv)$, global cost functions like $C_G(\thv)$ are untrainable for large problem sizes as they exhibit exponentially vanishing gradients (i.e., barren plateaus~\cite{mcclean2018barren}) even when the ansatz is short depth~\cite{cerezo2020cost}. Such barren plateaus can be avoided by employing a different type of cost function known as a local cost~\cite{cerezo2020cost,sharma2020}, where $C$ is defined such that one compares states or operators  with respect to each individual qubit rather than comparing them in a global sense.

One can construct a local cost where the Hamiltonian is a weighted sum of local $z$-Pauli operators:
\begin{equation}\label{eq:Hamlocal}
  C_L\equiv \langle H_L\rangle\,,\quad  H_L  = \id - \sum_{j=1}^n r_j Z_j\,,
\end{equation}
where $r_j \in \mathbb{R}$ and $Z_j$ is the $z$-Pauli operator acting on qubit $j$. Care must be taken when choosing the coefficients $\{r_j\}_{j=1}^n$ to ensure that the lowest $m$-eigenenergies of $H_L$ are non-degenerate. For instance,  when targeting the largest eigenvalue of $\rho$ ($m=1$),  the simple choice $r_j = 1$,  $\forall j$ achieves this goal. On the other hand, if one is interested in $m = n+1$ eigenvalues, then one can choose $r_j = r_1 + (j-1)\delta$ with $r_1 \gg \delta $, which will ensure that the $m$-lowest energy levels, $\{E_1, E_1 +r_1, E_1 +r_1 + \delta, ..., E_1  +r_1 + (m-1)\delta\}$, are non-degenerate. Henceforth, we will assume that one has chosen $\{r_j\}_{j=1}^n$  such that the $m$-lowest energy levels are non-degenerate.

While fixed local cost functions do not exhibit barren plateaus for shallow depth, they still have several trainability issues. First, having less degeneracy in $H_L$ leads to a more difficult optimization problem. Since degeneracy allows for additional freedom in the solution space, non-degeneracy constrains the possible solutions. Therefore, there is a tradeoff between engineering non-degeneracy (which allows one to distinguish more eigenvalues of $\rho$) versus keeping degeneracy (which allows for more solutions). Second, we expect both $C_L$ and $C_G$ to have a high density of local minima, especially for large $m$. This is because there will be partial solutions to the problem where one correctly assigns some eigenvalues of $\rho$ to the right energy levels of the Hamiltonian, while incorrectly assigning other eigenvalues. This local minima issue is what motivates the following adaptive approach.

\begin{figure}[t]
    \centering
    \includegraphics[width=.8\columnwidth]{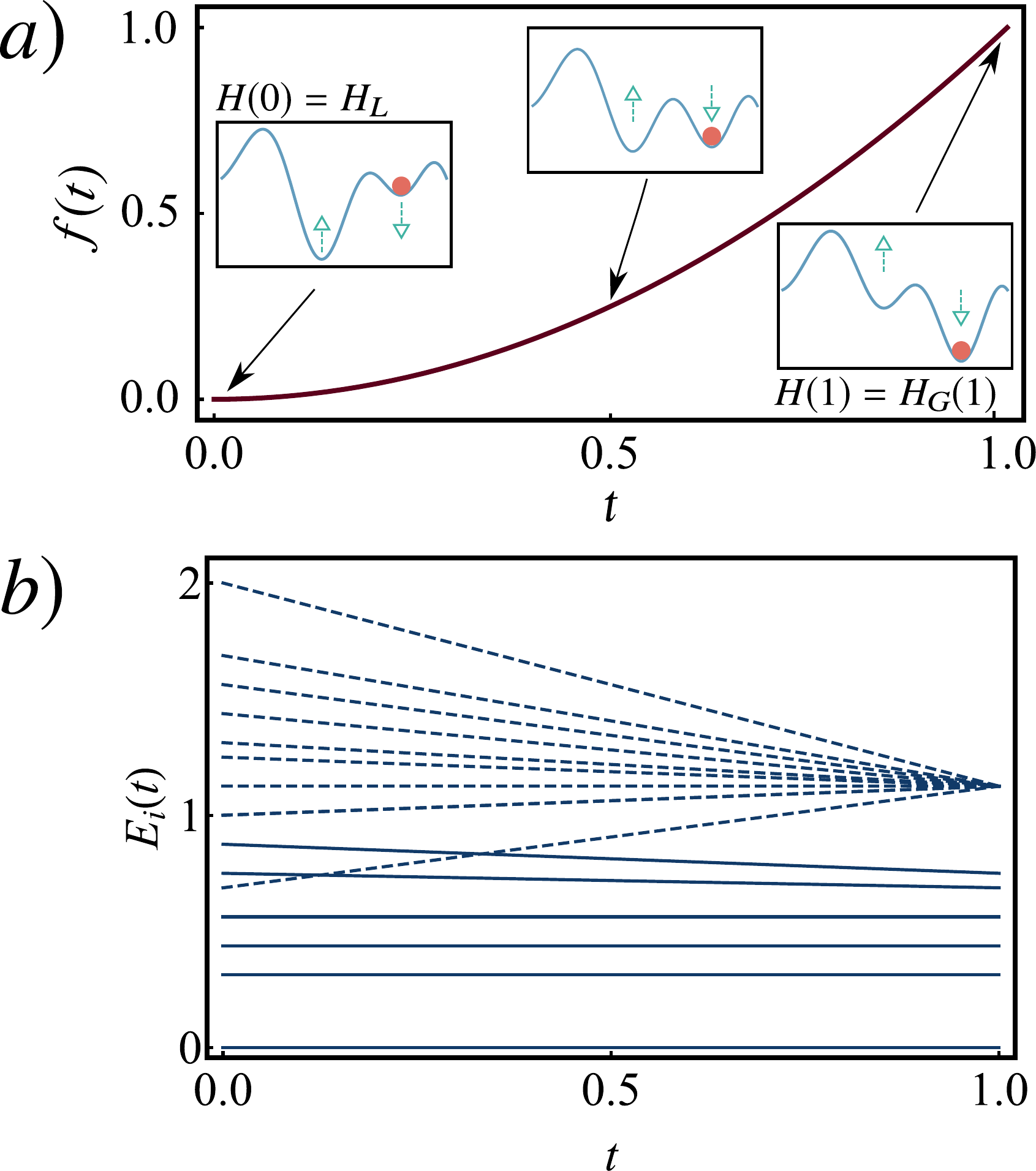}
    \caption{\textbf{Adaptive cost function}. (a) Schematic representation of the function $f(t)$ and the cost landscape of $C(t,\thv)$ versus $t$. We choose $f(t)$ as a slowly growing function with $t$. When the optimization starts at $t=0$, the cost function does not exhibit a barren plateau as the  Hamiltonian is local $H(0)=H_L$.  As $t$ increases $H(t)$ becomes a linear combination of $H_L$ and a global Hamiltonian $H_G(t)$ which is adaptively updated using information gained from measurements on $V(\thv)\rho V\ad(\thv)$. As shown in the insets, this procedure allows for local minima to become global minima. Finally, when the algorithm ends at $t=1$ the Hamiltonian is global  $H(t)=H_G(t)$. (b) Schematic representation of the eigenenergies of $H(t)$ versus $t$. For small $t$ the Hamiltonian is local and hence its spectrum contains non-degenerancies that reduce the space of solutions. At $t=1$, $H(t)$ becomes a global Hamiltonian and the spectrum has $m$ non-degenerate levels and a $(2^n-m)$-degenerate level.   
    }
    \label{fig:adaptive}
\end{figure}

\textbf{Adaptive Hamiltonian}. Let us now we  introduce an approach to adaptively update the VQSE Hamiltonian (and hence the cost function) based on information obtained via measurements during the optimization loop. This method allows us to mitigate the issues discussed in the previous section that arise for cost functions with fixed local or global Hamiltonians. Namely, the adaptive cost function solves the following three problems: (1) barren plateaus for shallow depth~\cite{cerezo2020cost}, (2) high density of local minima, (3) smaller solution space arising from non-degenerancies.

Consider a time-dependent Hamiltonian of the form
\begin{equation}
    H(t)\equiv (1-f(t))H_L+f(t) H_G(t)\,, 
\end{equation}
where $f(t)$ is a real-valued function such that $f(0)=0$, $f(1)=1$, and $H_L$ is a local Hamiltonian as in~\eqref{eq:Hamlocal}. We recall here that $t\in[0,1]$ is a parameter that indicates the optimization loop run time. Moreover, we define the time-dependent global Hamiltonian
\begin{equation}\label{eq:globaladaptive}
    H_G(t)\equiv\id-\sum_{i=1}^{m} q_i \ketbra{\zv_i(t)}{\zv_i(t)}\,,
\end{equation}
where the coefficients $q_i$ are real and positive, and chosen in the same way as in~\eqref{eq:costglobalm}. In addition, the states $\ket{\zv_i(t)}$  are adaptively chosen throughout the optimization loop by preparing $\widetilde{\rho}$, measuring in the standard basis to obtain the sets $\LC$ and $\ZC$, and updating $H_G(t)$ so that $\zv_i(t)\in\ZC$.

As schematically shown in Fig.~\ref{fig:adaptive}(a), in  order to mitigate the barren plateau phenomena it is important to choose a function $f(t)$ which is not rapidly growing with $t$. Hence, for small $t$, $H(t)\sim H_L$ and the cost function will be trainable as it will not present a barren plateau. Then, as $t$ increases, one can deal with the issue of local minima by updating $H_G(t)$. As depicted in the insets  of Fig.~\ref{fig:adaptive}(a), adaptively changing $H_G(t)$ transforms local minima in the cost landscape into global minima. Then,  by the end of the algorithm we have $H(1)= H_G(1)$, and as shown in panel (b) of Fig.~\ref{fig:adaptive}, the spectrum of $H$ becomes highly degenerate and the dimension of the solution space increases. In  Section~\ref{sec:alg} of the Methods  we present an algorithm to illustrate how one can update $H(t)$.

We remark that  Ref.~\cite{garcia2018addressing} proposed a method called adiabatically assisted VQE (AAVQE), which dynamically updates the VQE cost function by driving between a simple Hamiltonian to the non-trivial problem Hamiltonan. Note that  the goals of AAVQE and our adaptive training method are diffferent. Furthermore, in our method one adaptively updates the cost function based on information obtained through measurements, while AAVQE does not use information gained during the optimization.

\begin{figure}[t]
    \centering
    \includegraphics[width=\columnwidth]{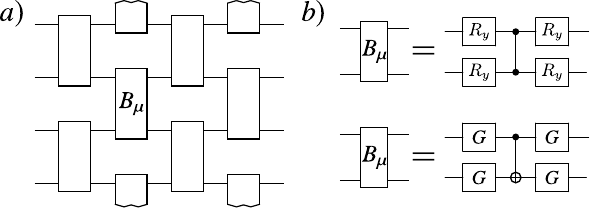}
\caption{ \textbf{Ansatz diagram.} (a) Layered hardware-efficient ansatz for $V(\thv)$. A single layer of the ansatz is composed of two-qubit gates $B_\mu(\thv_\mu)$ acting on neighboring qubits. Shown is the case of two layers. (b) While there are many choices for each block $B_\mu(\thv_\mu)$, in our numerics we employed two different parameterizations. Top: Each gate is composed of a controlled-$Z$ gate preceded and followed by single-qubit rotations about the $y$-axis $R_y(\theta)=e^{i\theta\sigma_y}$. Bottom: Each gate is composed of a CNOT gate preceded and followed by a single-qubit rotation $
G(\theta_1,\theta_2,\theta_3)= e^{i \theta_3 \sigma_z/2}e^{i \theta_2 \sigma_y/2}e^{i \theta_1 \sigma_z/2}$. The number of parameters in $\thv$ increases linearly with the number of layers and the number of qubits $n$.}
    \label{fig:ansatz}
\end{figure}

\textbf{Operational meaning of the cost function.} Here we discuss the operational meaning of the VQSE cost function, showing that small cost values imply small eigenvalue and eigenvector errors.  Let $\{\ket{\widetilde{\lambda}_i}\}_{i=1}^m$ be the set of the inferred eigenvector associated with every $\widetilde{\lambda}_i$ in $\LC$, and let $\ket{\delta_i}=\rho\ket{\widetilde{\lambda}_i}  -\widetilde{\lambda}_i\ket{\widetilde{\lambda}_i}$.  
We then define eigenvalue and eigenvector errors as follows:
\begin{equation}\label{eq:eig-error}
    \varepsilon_\lambda\equiv \sum_{i=1}^{m}(\lambda_i-\widetilde{\lambda}_i)^2, \quad \varepsilon_v\equiv \sum_{i=1}^{m}\braket{\delta_i}{\delta_i}.
\end{equation}
Here $\braket{\delta_i}{\delta_i}$ quantifies the component of $\rho \ket{\widetilde{\lambda}_i}$ that is orthogonal to $\ket{\widetilde{\lambda}_i}$, which follows from the following identity: $\ket{\delta_i} = (\id - \ket{\widetilde{\lambda}_i}\bra{\widetilde{\lambda}_i}) \rho \ket{\widetilde{\lambda}_i}$.

Then by using the Cauchy-Schwarz inequality, majorization conditions, and Schur convexity, we establish the following upper bound on eigenvalue and eigenvector errors (see Section \ref{sec:opermean} for more details): 
\begin{equation}\label{eq:opmeaningCErr}
   \varepsilon_\lambda, \varepsilon_v\leq \Tr[\rho^2]-\frac{(E_{m+1} - C(\thv))^2}{\sum_{i=1}^{m} (E_{m+1} - E_i)^2}\,,
\end{equation}
where $(E_1, \dots, E_m)$ are the $m$-smallest eigeneneries of $H$, and where for simplicity we have omitted the $t$ dependence. Thus Eq.~\eqref{eq:opmeaningCErr} provides an operational meaning to our cost function, as  small values of the cost function lead to small eigenvalue and eigenvector errors.

\begin{figure*}[ht]
    \centering
    \includegraphics[width=\linewidth]{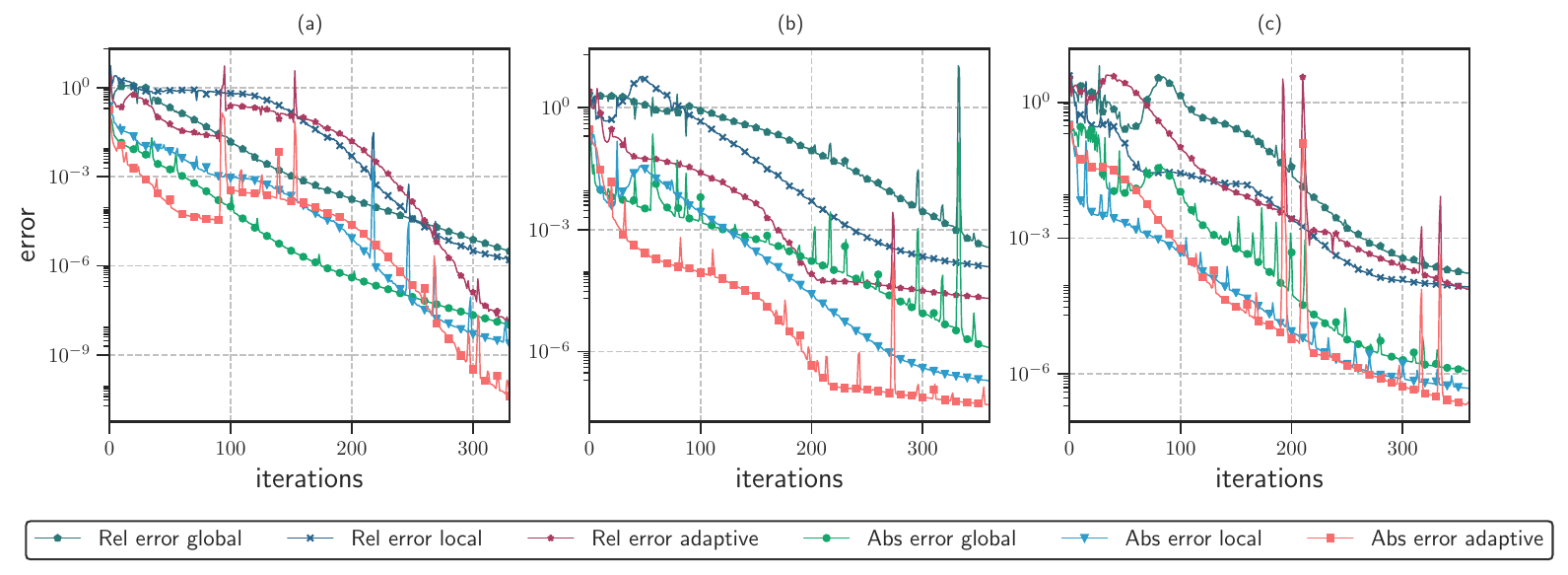}
    \caption{\textbf{Relative and absolute error versus the number of iterations}. We implemented VQSE  for states of: (a) $n=6$, (b)  $n=8$,  and (c) $n=10$ qubits. In all cases the ansatz for $V(\vec{\theta})$ was given by three layers of the Layered Hardware Efficient Ansatz of Fig.~\ref{fig:ansatz}(b, top).   Each curve represents the absolute or relative error (denoted Abs error or Rel error, respectively) of~\eqref{eq:rel-error} obtained by training $V(\vec{\theta})$ when employing an adaptive, fixed-local, or fixed-global Hamiltonian. The number of iterations was 330 for (a) and 360 for (b) and (c). For the adaptive runs we employed  Algorithm~\ref{alg1}, with  the Hamiltonian being updated every 30 iterations. 
    In each case the adaptive approach performs the best as it achieves the smallest errors.
 }
    \label{fig:error}
\end{figure*}

\subsection{Verification of solution quality}\label{sec:verif}

Let us show how to verify the results obtained from the VQSE algorithm. We remark that this verification step is optional, particularly because it requires $2n$ qubits, whereas the rest of VQSE only requires $n$ qubits.

In Section~\ref{sec:verifmeth} of Methods, we prove the following useful bound on eigenvalue and eigenvector error:
\begin{equation}\label{eq:verif}
    \varepsilon_\lambda, \varepsilon_v\leq \Tr[\rho^2] - \left(\sum_{i=1}^{\widehat{m}} \widetilde{\lambda}_i^2 +\frac{(1-\sum_{i=1}^{\widehat{m}} \widetilde{\lambda}_i)^2}{2^n-\widehat{m}}\right) \,,
\end{equation}
where one can take $\widehat{m}$ as any integer between $m$ and $2^n$. One can efficiently estimate the right-hand-side of \eqref{eq:verif} as follows. Given two copies of $\rho$, $\Tr[\rho^2]$ can be estimated by a depth-two quantum circuit with classical post-processing that scales linearly with  $n$~\cite{cincio2018learning}. Moreover, since $\Tr[\rho^2]$ is independent of $V(\thv)$, one only needs to compute it once (outside of the optimization loop). Estimating the $\widetilde{\lambda}_i$ for $i = 1,..., \widehat{m}$ essentially comes for free as part of the eigenvalue readout step of VQSE, where we note that taking $\widehat{m} > m$ simply involves keeping track of the frequencies of more bitstrings (more than the $m$-largest) during this readout step. Finally, we remark that while Eq.~\eqref{eq:opmeaningCErr} can also be used for  verification, in Section~\ref{sec:verifmeth} we show that~\eqref{eq:verif} provides a tighter bound, particularly as one increases $\widehat{m}$.

\subsection{Ansatz}

While there are many possible choices for the ansatz $V(\thv)$, we are here restricted to state-agnostic ansatzes which do not require any a prior information about~$\rho$. One such ansatz is the Layered Hardware Efficient Ansatz~\cite{kandala2017} shown in Fig.~\ref{fig:ansatz}(a). Here, $V(\thv)$ consists of a fixed number  $L$ of layers of two-qubit gates $B_\mu(\thv_\mu)$ acting on alternating pairs of neighboring qubits. Figure~\ref{fig:ansatz}(b) illustrates possible choices for $B_\mu(\thv_\mu)$. Note that with this structure, the number of parameters in $\thv$ grows linearly  with the $n$ and $L$. 


Let us remark that the Layered Hardware Efficient Ansatz can lead to trainability issues as the system size increases~\cite{mcclean2018barren,cerezo2020cost}. Hence, different strategies have been proposed to mitigate such difficulties, such as learning to initialize parameters~\cite{verdon2019learning},  layer-by-layer training~\cite{grant2019initialization}, and correlating the parameters~\cite{volkoff2020large}. In addition, these methods can be combined with more sophisticated ansatzes, such as a variable-structure ansatz~\cite{larose2018,cincio2018learning} where the structure of the ansatz is not fixed, and where the gate placement becomes an optimizable hyper-parameter. This variable-structure approach has already been shown to improve performance in the context of extracting the eigensystem of a quantum state~\cite{larose2018}.

Finally, since VQSE optimization corresponds to an energy minimization problem, a natural ansatz that can also be used to mitigate trainability issues is the Quantum Alternating Operator Ansatz (QAOA)~\cite{farhi2014QAOA,hadfield2019quantum}. Specifically, one could employ $H(t)$ as the problem Hamiltonian in the QAOA and use a standard mixing Hamiltonian. While we do not employ this ansatz in our heuristics, it is nevertheless of interest for future work.

\subsection{Optimization}

Regarding the optimization of the parameters $\thv$, while gradient-free methods are an option~\cite{nakanishi2019,parrish2019}, there has been recent evidence that gradient-based methods can perform better~\cite{harrow2019,kubler2020adaptive,sweke2019stochastic,arrasmith2020operator}.  Moreover, as shown in~\cite{mitarai2018quantum,schuld2019evaluating},  for cost functions like~\eqref{eq:cost}, gradients can be analytically determined (see Section~\ref{sec:grad} in the Methods for an explicit derivation of the gradient formula). Therefore, in our heuristics, we employ gradient-based optimization.

\begin{figure*}[th]
    \centering
    \includegraphics[width=\linewidth]{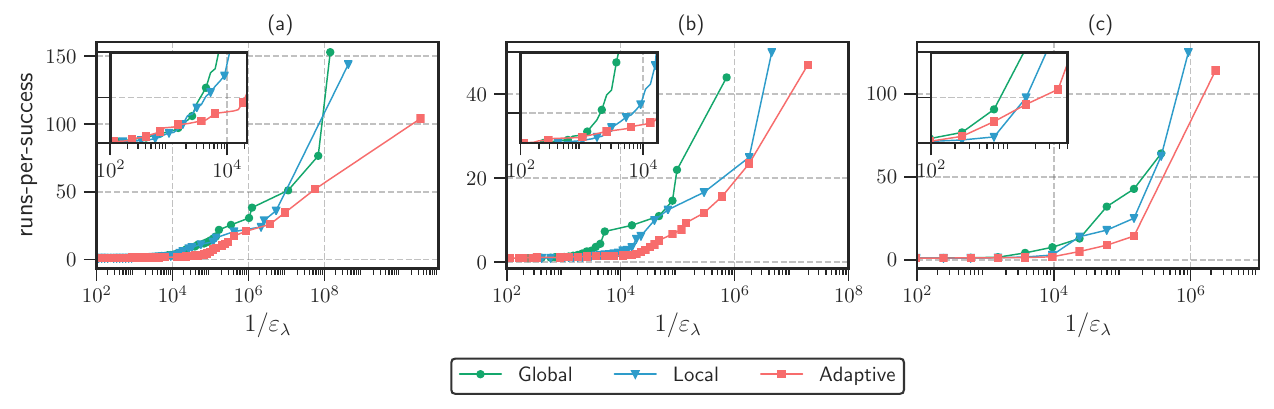}
    \caption{\textbf{Runs-per-success versus inverse absolute error $1/\varepsilon_{\lambda}$}.   We implemented VQSE  for the states of: (a) $n=6$, (b)  $n=8$,  and (c) $n=10$ qubits  corresponding to Fig. \ref{fig:error}. The insets depict the same data in the small $1/\varepsilon_{\lambda}$ regime.  Runs-per-success is defined as the total number of runs divided by the number of runs with a relative error smaller than a target $\varepsilon_{\lambda}$. 
    For all three cases, we can see that as $1/\varepsilon_{\lambda}$ increases, the adaptive Hamiltonian has the lowest number of runs-per-success, and hence the best performance. 
    In all cases the $x$ axis is plotted on a log scale.}
    \label{fig:rps}
\end{figure*}

\subsection{Numerical Implementations}

Here we present the numerical results obtained from implementing VQSE. We first employ VQSE to estimate the spectrum of quantum states of different dimensions and compare the performance of cost functions based on the global, local, and adaptive Hamiltonians  discussed in Section~\ref{sec:cost-funtions}. Then we use VQSE for error mitigation of the $W$-state preparation circuit. Finally, we implement VQSE for entanglement spectroscopy on the ground-state of an $XY$-spin chain, which allows us to detect the presence of quantum critical points.

\textbf{VQSE for quantum principal component analysis.} Figure~\ref{fig:error} presents the results of implementing VQSE to estimate the  six largest eigenvalues ($m=6$) of quantum states with $n=6,8,$ and $10$ qubits. In all  cases we have $\rank(\rho)=16$, as the states were prepared by randomly entangling the system qubits with four ancillary qubits, which were later traced out. Moreover, we chose $\rho$ to be real and not sparse in the standard basis. 


In our heuristics we used the Layered Hardware Efficient Ansatz of Fig.~\ref{fig:ansatz}(b, top), and we employed the  fixed-local, fixed-global, and adaptive cost functions of Section~\ref{sec:cost-funtions}.  The termination condition was stated in terms of the maximum number of iterations in the optimization loop. Hardware noise and finite sampling were not included in these heuristics. (The next subsection shows heuristics with noise.) For the fixed local cost function we chose the $\{r_j\}_{j=1}^n$ in \eqref{eq:Hamlocal} so that the first six energy eigenvalues of $H_L$ were non-degenerate. Moreover, we defined the fixed global Hamiltonian such that the first six energy levels (i.e.,  associated eigenvectors and spectral gaps) coincided with those of $H_L$.  Finally, the adaptive Hamiltonian was constructed according to the procedure described above, and more specifically, in Algorithm~\ref{alg1} in the Methods section.

Since for these examples we can calculate the exact eigenvalues $\lambda_i$,  we compute and plot the following quantities which we use as figures of merit for the performance of the VQSE algorithm: 
\begin{align}
    \varepsilon_{\lambda} \equiv \sum_{i=1}^6 (\lambda_i - \widetilde{\lambda}_i)^2\,, \quad
    \varepsilon_r \equiv \sum_{i=1}^6 (\lambda_i - \widetilde{\lambda}_i)^2/\lambda_i^2\,. \label{eq:rel-error}
\end{align}
Here $\varepsilon_{\lambda}$ and $\varepsilon_r$ respectively quantify absolute error and relative error in estimating the exact eigenvalues. We remark that these two quantities provide different information: The absolute error is biased towards the error in estimating the large eigenvalues of $\rho$, while on the other hand, the relative error is more sensitive to errors in estimating the small eigenvalues of $\rho$. 

Figure~\ref{fig:error} plots the relative and absolute errors versus number of iterations (with the total number of iterations fixed). While we performed many runs, these plots show only the run that achieved the lowest absolute error. For all system sizes considered ($n=6,8$ and $10$),   VQSE achieves smaller relative and absolute errors when employing the adaptive Hamiltonian approach than using a fixed Hamiltonian. For $n=6$, the errors obtained by adaptively updating $H(t)$ are two orders of magnitude smaller than those obtained with fixed Hamiltonians, while for $n=8$ they are one order of magnitude smaller.  As shown in Fig.~\ref{fig:error}(c) for $n=10$, the adaptive Hamiltonian approach achieves error of the order: $\sim 10^{-5}$ for the relative error, and $\sim 10^{-7}$ for the absolute error, and again outperforms the fixed local Hamiltonian approaches. We here finally remark that that  we can use Eq.~\eqref{eq:runsestimates} to  determine the number of shots needed to guarantee that with a probability larger that $99\%$ the relative error induced by finite sampling is smaller that  $0.001$. Namely, we find that one needs a number of shots larger than $50.2K$, which is well within the order of magnitude of shots regularly used.

It is natural to ask whether the runs shown in Fig.~\ref{fig:error} are representative of the algorithm performance. To provide an analysis of the average VQSE performance, we plot in Fig.~\ref{fig:rps}  the runs-per-success versus $1/\varepsilon_{\lambda}$ for each of the aforementioned examples.  Here, runs-per-success is defined as the total number of runs divided by the number of runs with an absolute error smaller than a target~$\varepsilon_{\lambda}$.

From all three panels in  Fig.~\ref{fig:rps} we see  that for large $1/\varepsilon_{\lambda}$, the adaptive Hamiltonian always has the best performance as it requires less runs-per-success to achieve smaller errors. Finally, it is interesting to note from the insets of Fig.~\ref{fig:rps} that there is a  regime where   the run time has a linear dependence on $\log (1/\varepsilon_{\lambda})$ when employing an adaptive approach. This suggests that VQSE may perform quite efficiently for large $\varepsilon_{\lambda}$. However, the linear dependence breaks down for small $\varepsilon_{\lambda}$, where the number of runs-per-success seems to grow exponential with $1/\varepsilon_{\lambda}$. Despite, such growth, for up to $8$ qubits we only need 100 repetitions to achieve an error of order $10^{-6}$. We leave for future work a more detailed study of the dependence of runs-per-success for small error. Finally, the results presented in Figs.~\ref{fig:error} and~\ref{fig:rps} suggest that for a sufficiently large value of number of iterations, the adaptive Hamiltonian approach outperforms the fixed Hamiltonian approaches as it requires the least number of iterations to converge to very small values of $\varepsilon_{\lambda}$.

\begin{figure}[t]
     \centering
    \includegraphics[width=0.9\columnwidth]{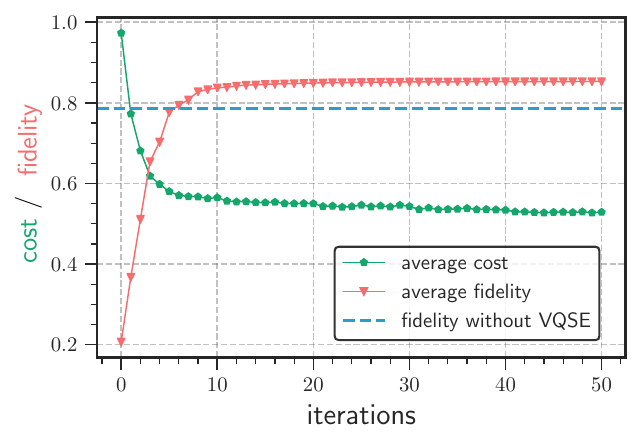}
    \caption{ \textbf{Cost function value and fidelity versus  number of iterations}. We implement 
    VQSE for error mitigation of the three qubit $W$-state preparation circuit. The input state $\rho$ corresponds to the mixed state obtained by running the $W$-state preparation circuit on a noisy simulator. The dashed line corresponds to the fidelity $F(\rho,\ket{\psi})$ between $\rho$ and the the exact $W$ state $\ket{\psi}$.
    For each iteration step, we compute the fidelity $F(\sigma,\ket{\psi})$, where the mixed state $\sigma$ is obtained by running the VQSE eigenvector preparation circuit on the noisy simulator. 
    Curves depict the average of 10 instances of the algorithm.  As the number of iterations increases the cost function value decreases, which implies that we are able to train $V(\thv)$ in the presence of noise. After a few iterations of the VQSE optimization loop, we find $F(\sigma,\ket{\psi})> F(\rho,\ket{\psi})$.}
    \label{fig:em}
\end{figure}

\textbf{Error mitigation.} Here we discuss an important application of the VQSE algorithm for error mitigation. Quantum state preparation circuits (gate sequences $U$ which prepare a target state $\ket{\psi}$) are used as subroutines in many quantum  algorithms. However, since current quantum computers are noisy, all state preparation circuits produce mixed states $\rho$. If there is little enough incoherent noise, we can expect that the largest eigenvalue of $\rho$ is associated with $\ket{\psi}$. Here we show that VQSE can be implemented to re-purify $\rho$  and estimate $\ket{\psi}$. Naturally, when running the VQSE eigenvector preparation circuit, noise will also produce a mixed state $\sigma$. However, if the depth of $V(\thv)$ is shorter than the depth of $U$, one can obtain a higher fidelity between $\sigma$ and $\ket{\psi}$ in comparison to the fidelity between $\rho$ and $\ket{\psi}$. In this case one can 
mitigate errors by  replacing the state preparation circuit  by the VQSE eigenvector preparation circuit.

Let us now consider the three qubit $W$-state preparation circuit from \cite[Section~2.2]{cruz2019efficient} (see also \cite{bartschi2019deterministic}). By employing a noisy quantum computer simulator with the  noise profile of IBM's Melbourne
processor~\cite{IBMQ14}, we find that the fidelity between $\rho$ and the exact $W$ state $\ket{\psi}$ is  $F(\rho,\ket{\psi})\approx 0.785$. We then train 10 instances of VQSE with two layers of the ansatz in Fig.~\ref{fig:ansatz}(b, bottom) and with a termination condition of 50 iterations. Moreover, we employ the adaptive Hamiltonian, where we update $H(t)$ every $10$ iterations according to Algorithm~\ref{alg1}. Figure~\ref{fig:em} shows the average cost function value and  average fidelity between $\ket{\psi}$ and the state $\sigma$ obtained by running the VQSE eigenvector preparation circuit. As the number of iterations increases, the cost value tends to decrease, showing that we are able to train in the presence of noise. Moreover,  we also see that  $F(\sigma,\ket{\psi})$ increases and saturates at a value larger than  $F(\rho,\ket{\psi})$, namely at 0.853, hence showing that we are in fact mitigating the effect of noise. This can be explained by the fact that we reduced the circuit depth, as  our ansatz contains two CNOTs, while the textbook circuit contains three CNOTs.

\textbf{Entanglement Spectroscopy}. We now discuss the possibility of employing VQSE  to compute the entanglement spectrum of a state $\rho$ which is obtained  as the reduced state of a bipartite quantum system $\ket{\psi_{AB}}$, i.e., $\rho=\Tr_B \dya{\psi_{AB}}$. Let $d$ denote the dimension of $\rho$. The entanglement spectrum~\cite{HaldaneES} refers to the collection $\{\lambda_k\}_{k=1}^d$ of eigenvalues  of $\rho$, and as discussed in~\cite{Suba__2019}, entanglement spectroscopy is a useful tool to analyze states $\ket{\psi_{AB}}$ prepared by simulating many-body systems on a quantum computer. Specifically, the entanglement spectrum is useful to study the bipartite entanglement, as it contains more universal signatures than the von Neumann entropy alone~\cite{HaldaneES}, and it can detect the presence of quantum critical points~\cite{GiampaES,cerezo2015}.

Let us now consider an $N=8$ spin-$1/2$ cyclic chain interacting trough uniform $XY$ first-neighbor Heisenberg coupling in the presence of a non-transverse magnetic field. The Hamiltonian of the system is
\begin{equation} \label{eq:H-spins}
H=-\sum_j(h_x S^x_j+h_z S^z_j+J_x S^x_jS^x_{j+1}+J_y S^y_jS^y_{j+1})\,,
\end{equation}
where $j$ labels the site in the chain, $S_j^\mu$ the spin operator (with $\mu=x,y,z$), $J_\mu$ the coupling strength, and $h_\mu$ the magnetic fields. Here, $J_\mu>0$ leads to ferromagnetic (FM) coupling, while  $J_\mu<0$ to antiferromagnetic (AFM) coupling.  As shown in~\cite{cerezo2015,cerezo2016}, for specific values of the fields $h_\mu$ (known as {\it factorizing fields}) the Hamiltonian in~\eqref{eq:H-spins} presents quantum critical points known as ``factorization'' points.  At the non-transverse factorizing field, the ground-state of $H$ becomes a separable non-degenerate state such that one of its eigenvalues is exactly equal to one, while the rest are exactly zero.

\begin{figure}[t]
    \centering
    \includegraphics[width=\columnwidth]{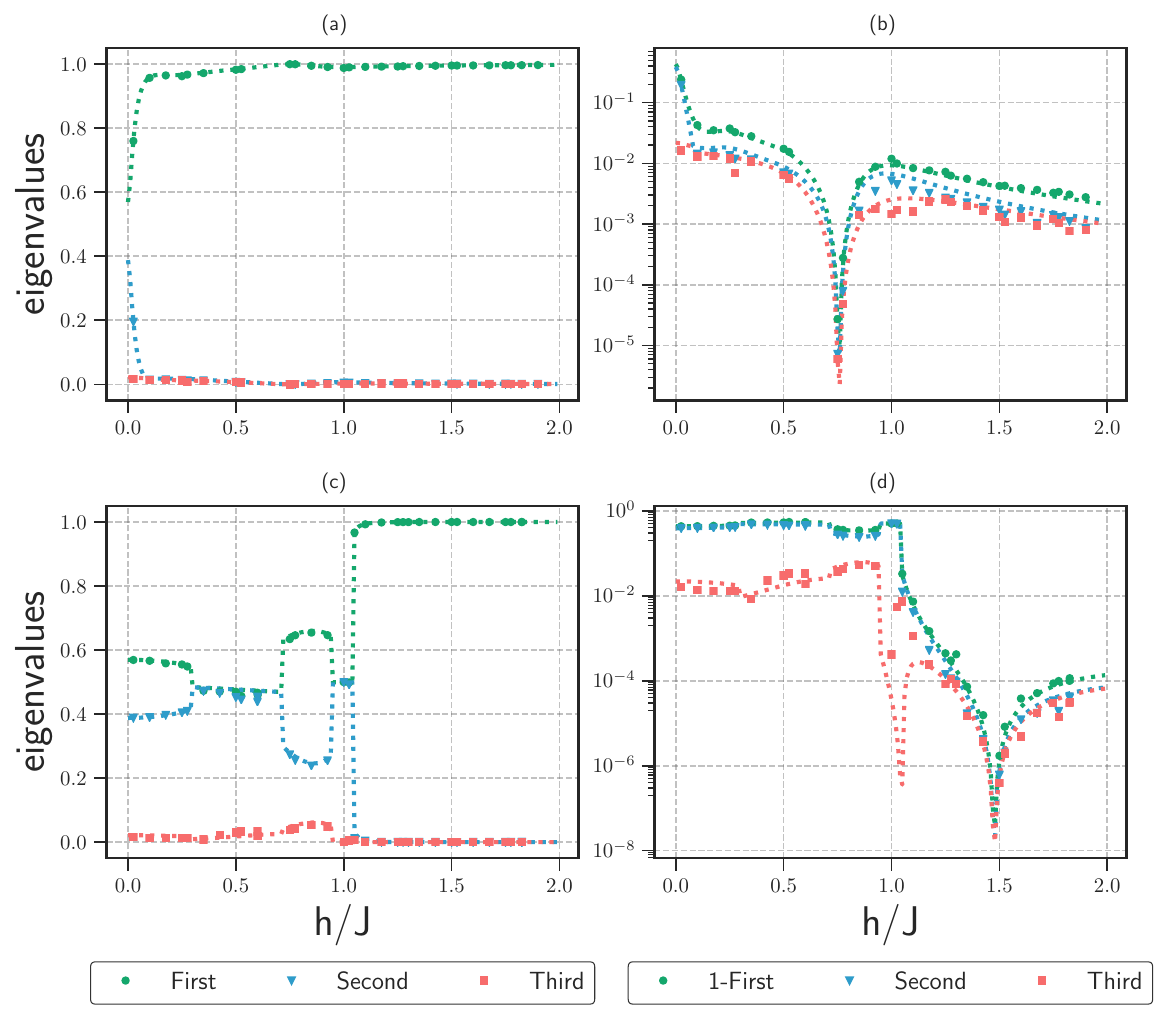}
    \caption{\textbf{Exact and estimated eigenvalues versus field value, for the VQSE  entanglement spectroscopy  implementations}. The input state $\rho$ is given as the reduced state of 4 neighboring qubits from the ground state of~\eqref{eq:H-spins}. Top and bottom rows correspond to ferromagnetic and antiferromagnetic couplings, respectively. Dashed curves represent the exact three largest eigenvalues of $\rho$, while plot markers indicate the VQSE estimated eigenvalues. In (a) and (c) we see that VQSE can accurately estimate the eigenvalues.   In (b) and (d) we plot $1-\lambda_1$, and the $y$ axis is on a log scale. Here we see that the quantum critical factorization points are detected at $h/J_x \approx 0.76$ and $h/J_x \approx 1.43$ in (b) and (d),  respectively, since at those points we have  $\widetilde{\lambda}_1\approx1$, and $\widetilde{\lambda}_2,\widetilde{\lambda}_3\approx0$.}
    \label{fig:spin}
\end{figure}

In Fig.~\ref{fig:spin}(a) and (c), we show results of implementing VQSE with an adaptive Hamiltonian to compute the three largest eigenvalues of the state $\rho$ defined as the reduced state of $4$ neighboring spins obtained from the ground state of~\eqref{eq:H-spins}. For simplicity we have parametrized the fields as $(h_z,h_x)=h(\cos(\gamma),\sin(\gamma))$ with $\gamma$ fixed. Specifically, in Fig. \ref{fig:spin}(a) and (c) we plot the estimated eigenvalues versus the field magnitude $h$ for a system with FM and AFM couplings, respectively. Moreover, dashed lines indicate the exact eigenvalues. For each field value, we run $8$ instances of  VQSE, and even for such a small number of runs, the estimated eigenvalues give good approximations  as we get relative errors which in general are of the order of $\sim10^{-2}$.

In Fig.~\ref{fig:spin}(b) and (d), we show the same data as in (a) and (c) but the $y$ axis is plotted on a logarithmic scale, and where instead of plotting the largest eigenvalue $\lambda_1$, we plot $1-\lambda_1$. For the FM (AFM) case, there is a factorization points at $h/J_x\approx 0.76$ ($h/J_x\approx 1.43$). As depicted in these panels, around critical points we correctly find  $\widetilde{\lambda}_1\approx 1$, and  $\widetilde{\lambda}_2, \widetilde{\lambda}_3\approx 0$. These results show that VQSE can detect  quantum critical factorization points.

\section{Discussion}

In the NISQ era, every qubit and every gate counts. Wasteful usage of qubits or gates will ultimately limit the problem size that an algorithm can solve. In this work, we presented an algorithm for extracting the eigensystem of a quantum state $\rho$ that is as frugal as we could imagine, with respect to qubit count.  

We introduced the Variational Quantum State Eigensolver (VQSE), which estimates the $m$-largest eigenvalues and associated eigenvectors of $\rho$, using only a single copy of $\rho$, and hence only $n$ qubits per iteration of the VQSE. VQSE exploits the mathematical connection between diagonalization and majorization to define an efficiently computable cost function as the expectation value of a Hamiltonian. We derived an operational meaning of this cost function as a bound on eigensystem error. Furthermore, we introduced a  training method that involved adaptively updating the VQSE cost function based on the information gained from measurements performed throughout the optimization. This was aimed at addressing both barren plateaus and local minima in the cost landscape. 

We have numerically implemented VQSE for several applications. We showed that VQSE can be employed for PCA by implementing the VQSE algorithm on states of $n=6,8,$ and $10$ qubits to estimate the six largest eigenvalues. Our numerical results (Figs.~\ref{fig:error} and \ref{fig:rps}) indicate that our adaptive cost function approach leads to smaller errors than the ones obtained by training a fixed cost function.  We also showed (Fig.~\ref{fig:spin}) that one can detect quantum critical points by performing entanglement spectroscopy with the eigenvalues obtained via VQSE. Finally, we employed VQSE to mitigate errors that occur during the $W$-state preparation circuit. This involved running VQSE on a noisy simulator to re-purify the state, i.e., find the circuit that prepares the eigenvector with the largest eigenvalue. We found (Fig.~\ref{fig:em}) that the re-purified state obtained by VQSE improved the fidelity with the target $W$ state, and hence reduced the effects of noise.

\textbf{Comparison to literature.} Since VQSE only requires $n$ qubits, it is as qubit frugal as it can possible be when compared to other algorithms for the same task, such as quantum Principal Component Analysis (qPCA)~\cite{lloyd2014quantum},  Variational Quantum State Diagonalization (VQSD)~\cite{larose2018}, and Quantum State Singular Value Decomposition (QSVD)~\cite{bravo2019quantum}.  The quantum phase estimation and density matrix exponentiation primitives in qPCA make it difficult to implement in the near term~\cite{nielsen2010}, and this is supported an by attempted implementation in \cite{larose2018} that resulted in poor performance. On the other hand, VQSD and QSVD are variational algorithms and hence have the possibility of lower-depth requirements. But they still need to employ a larger number of qubits than VQSE. Specifically, VQSD needs to perform the so-called Diagonalized Inner Product Test~\cite{larose2018} that requires two copies of $\rho$, i.e., requires twice as many qubits as VQSE. In addition, it is also worth noting that VQSD is vulnerable to noise, since any asymmetry between the noise acting of each copy of $\rho$ will affect the result of the algorithm. Finally, in QSVD, one needs to either compute or have access to a purification $\ket{\psi}$ of $\rho$. Hence QSVD requires a number of qubits between $n$ and $2n$. Moreover, we expect that noise will be a bigger issue for QSVD than for VQSE, since in practice the assumption that one has a pure state in QSVD can often be violated due to incoherent noise during state preparation.

Quantum-inspired classical algorithms~\cite{tang2019quantum} for PCA can  perform well in practice, provided that the matrix has a very large dimension, low rank, and low condition number~\cite{arrazola2019quantum}. We note that VQSE does not have such limitations, except the fact that VQSE yields results with high accuracy for low-rank states. Here is is also paramount to recall that recent results have show that an exponential advantage is still possible for PCA~\cite{cotler2021revisiting,huang2021quantumadvantage}, even for near-term algorithms, as the quantum-inspired classical algorithms are artificially given too much power via access to quantum state amplitudes. Thus, in view of these recent result, VQSE can be useful in the quest for achieving  a quantum speedup with quantum PCA, particularly for analysis of quantum data. For the case of classical data, the success of VQSE for PCA relies on the efficiency of preparing a quantum state corresponding to the covariance matrix of the classical data~\cite{aaronson2015read}. In addition, VQSE has applications not only for PCA but also for other tasks such as entanglement spectroscopy and error mitigation on NISQ devices. For error mitigation, we leave for future work combining our approach with other error mitigation techniques such as Virtual Distillation~\cite{koczor2020exponential,huggins2020virtual}, which also seeks to re-purify noisy quantum states.

\textbf{Future Directions.} Due to the rapid rise of VQE~\cite{peruzzo2014VQE}, much research has gone into how to prepare ground and excited states on NISQ devices. However, more research is needed on how to characterize these states, once prepared. This is where VQSE comes in, as VQSE can extract the entanglement spectra of these states and hence characterize important properties like topological order~\cite{HaldaneES}. Hence it is worth exploring in the future the idea of pairing up the VQE and VQSE algorithms, where VQSE is implemented immediately after VQE.

Furthermore, VQSE has immediate application for estimating the fidelity of two quantum states with reduced resource requirements. This is because an algorithm was previously introduced~\cite{cerezo2019variational} to estimate fidelity by using state diagonlization as a subroutine, and hence VQSE can provide a more efficient version of this subroutine. 

A crucial technical idea in this work was our adaptively-updated cost function, which improved optimization performance. It is worth investigating whether this adaptive method can improve the performance of other variational quantum algorithms~\cite{peruzzo2014VQE, farhi2014QAOA, johnson2017qvector, romero2017quantum, larose2018, arrasmith2019variational, cerezo2019variational, jones2019variational, yuan2018theory, li2017efficient, kokail2019self, Khatri2019quantumassisted, jones2018quantum, heya2018variational, endo2018variational,sharma2019noise, carolan2019variational,yoshioka2019variational,bravo-prieto2019,xu2019variational,mcardle2019variational,cirstoiu2019variational,otten2019noise,LubaschVariational20,verdon2019quantum,bravo2019quantum}. 

Another direction to explore is whether VQSE exhibits noise resilience~\cite{sharma2019noise}. We suspect this to be true given the similar structure of VQSE and the variational quantum compiling algorithms investigated in Ref.~\cite{sharma2019noise}.

This is important as we are proposing that VQSE will be a useful tool for error mitigation. Namely, we envision that VQSE could be used as a subroutine to improve the accuracy of several quantum algorithms. For example, one could use VQSE to re-purify the noisy quantum state obtained as the outcome of the VQE algorithm. Alternatively, one could periodically perform VQSE whilst running a dynamical quantum simulation on a NISQ device, which would re-purify the state as it is evolving in time. This could allow one to simulate long-time dynamics, i.e., times significantly beyond the coherence time of a NISQ device.

\section{Methods}

\subsection{Operational meaning of the cost function}\label{sec:opermean}
In this section, we provide a derivation for Eq.~\eqref{eq:opmeaningCErr}. First, we rewrite the eigenvalue error in Eq.~\eqref{eq:eig-error} as follows: 
\begin{align}\label{eq:eig-error2}
    \varepsilon_\lambda=\vec{\lambda}^m\cdot\vec{\lambda}^m+\vec{\widetilde{\lambda}}^m\cdot\vec{\widetilde{\lambda}}^m-2\vec{\lambda}^m\cdot\vec{\widetilde{\lambda}}^m\,,
\end{align}
where $\vec{\lambda}^m\equiv(\lambda_1,\ldots,\lambda_m)$ and $\vec{\widetilde{\lambda}}^m\equiv (\widetilde{\lambda}_1,\ldots,\widetilde{\lambda}_m)$. Since the eigenvalues of a positive semidefinite  operator majorize its diagonal elements, we have that $\vec{\lambda}^m\succ\vec{\widetilde{\lambda}}^m$. Moreover, from the Schur convexity property of the dot product with an ordered vector, it follows that $\vec{\lambda}^m\cdot\vec{\widetilde{\lambda}}^m \geq\vec{\widetilde{\lambda}}^m\cdot\vec{\widetilde{\lambda}}^m$, which further implies the following inequality: 
\begin{equation}\label{eq:ineqel}
    \varepsilon_\lambda\leq  \vec{\lambda}^m\cdot\vec{\lambda}^m-\vec{\widetilde{\lambda}}^m\cdot\vec{\widetilde{\lambda}}^m\,.
\end{equation}
Similarly, from Eq.~\eqref{eq:eig-error} we get 
\begin{align}
    \varepsilon_v\leq\vec{\lambda}^m\cdot\vec{\lambda}^m-\vec{\widetilde{\lambda}}^m\cdot\vec{\widetilde{\lambda}}^m\,,\label{eq:eigenvecerr3}
\end{align}
where we again used the fact that the eigenvalues of a positive semidefinite operator majorize its diagonal elements, and hence $\vec{\lambda}^m\cdot\vec{\lambda}^m\geq \sum_{i=1}^{m}\matl{\widetilde{\lambda}_i}{\rho^2}{\widetilde{\lambda}_i}$\,.

We recall from Eq.~\eqref{eq:mincost} that the VQSE cost function can be expressed as $C=\sum_{i=1}^{d} E_i p_i$, where we omit the $\thv$, and $t$ dependence of $C$. Therefore, the following chain of inequalities hold:
\begin{align} 
    C &\geq \sum_{i=1}^{m} E_i p_i + E_{m+1}\sum_{i>m}  p_i\nonumber\\
    & = E_{m+1} - \bigg(\sum_{i=1}^{m} p_i(E_{m+1}-E_i)\bigg)\nonumber\\
    & \geq E_{m+1} - \sqrt{\bigg(\sum_{i=1}^{m}p_i^2\bigg) \bigg(\sum_{i=1}^{m}(E_{m+1}-E_i)^2\bigg)}, \label{eq:ineq2}
\end{align}
where $d=2^n$. The first inequality follows the fact that $E_i\geq E_{m+1}$, $\forall i\geq m+1$ and $\sum_{i>m}p_i=1-\sum_{i=1}^m p_i$. The second inequality follows from the Cauchy-Schwarz inequality for the dot product of two vectors $|\vec{u}\cdot\vec{v}|\leq|\vec{u}||\vec{v}|$. By combining Eq.~\eqref{eq:ineq2} with the fact  that $\sum_{i=1}^{m}\widetilde{\lambda}_i^2 \geq \sum_{i=1}^{m}p_i^2$ (since $\widetilde{\lambda}_i\in\LC$ are the largest diagonal elements of $\widetilde{\rho}$), we find that
\begin{equation}\label{eq:opmeaningC2}
    \sqrt{\sum_{i=1}^{m}\tilde{\lambda}_i^2} \geq \frac{E_{m+1} - C}{\sqrt{\sum_{i=1}^{m} (E_{m+1} - E_i)^2}}\,.
\end{equation}

Using the fact that $\vec{\lambda}^m\cdot\vec{\lambda}^m\leq \vec{\lambda}\cdot\vec{\lambda} =\Tr[\rho^2]$, we obtain the following equality from~\eqref{eq:opmeaningC2} 
\begin{align}
   \vec{\lambda}^m\cdot\vec{\lambda}^m -\vec{\widetilde{\lambda}}^m\cdot\vec{\widetilde{\lambda}}^m 
  \leq \Tr[\rho^2] -\frac{(E_{m+1} - C)^2}{\sum_{i=1}^{m} (E_{m+1} - E_i)^2}\,.\nonumber
\end{align}
Combining this with~\eqref{eq:ineqel} and \eqref{eq:eigenvecerr3} leads to~\eqref{eq:opmeaningCErr}.

\subsection{Verification of solution quality}\label{sec:verifmeth}

Here we provide a proof of Eq.~\eqref{eq:verif}, and we show that this bound is tighter than the bound in~\eqref{eq:opmeaningCErr}. From the definition of the eigenvalue and eigenvector error in~\eqref{eq:eig-error}, 
it is straightforward to see that $
    \varepsilon_\lambda\leq \sum_{i=1}^{d}(\lambda_i-\widetilde{\lambda}_i)^2$, and $ \varepsilon_v\leq \sum_{i=1}^{d}\braket{\delta_i}{\delta_i}$,
where $d=2^n$. By following a procedure similar to the one employed in deriving~\eqref{eq:eigenvecerr3}, we  find 
\begin{equation}\label{eq:eig-error5}
    \varepsilon_\lambda\leq \vec{\lambda}\cdot\vec{\lambda}-\vec{\widetilde{\lambda}}\cdot\vec{\widetilde{\lambda}}\,,
\end{equation}
where we recall that $\vec{\lambda}$ and  $\vec{\widetilde{\lambda}}$ denote $d$-dimensional vectors of ordered exact and estimated eigenvalues of $\rho$, respectively. 
Moreover, from $\ket{\delta_i} = (\id - \ket{\widetilde{\lambda}_i}\bra{\widetilde{\lambda}_i}) \rho \ket{\widetilde{\lambda}_i}$, it is straightforward to get
\begin{align}
    \varepsilon_v\leq\sum_{i=1}^{d}\matl{\widetilde{\lambda}_i}{\rho^2}{\widetilde{\lambda}_i}-\vec{\widetilde{\lambda}}\cdot\vec{\widetilde{\lambda}} = \vec{\lambda}\cdot\vec{\lambda}-\vec{\widetilde{\lambda}}\cdot\vec{\widetilde{\lambda}}\,,\label{eq:eigenvecerr2}
\end{align}
where we used the fact that $\sum_{i=1}^{d}\matl{\widetilde{\lambda}_i}{\rho^2}{\widetilde{\lambda}_i}=\Tr[\rho^2]$, which follows from the invariance of trace under a basis transformation. 

Let $\vec{\widehat{\lambda}}=(\widetilde{\lambda}_1,\ldots,\widetilde{\lambda}_{\widehat{m}},\frac{1-\sum_{i=1}^{\widehat{m}} \widetilde{\lambda}_i}{2^n-\widehat{m}},\ldots,\frac{1-\sum_{i=1}^{\widehat{m}} \widetilde{\lambda}_i}{2^n-\widehat{m}})$, with $\widehat{m}>m$, be a vector  majorized by $\vec{\widetilde{\lambda}}$, i.e.,   $\vec{\widetilde{\lambda}}\succ\vec{\widehat{\lambda}}$. Since  the dot product with an ordered vector is a Schur convex function, we have  $\vec{\widehat{\lambda}}\cdot \vec{\widehat{\lambda}}\leq \vec{\widehat{\lambda}}\cdot \vec{\widetilde{\lambda}}\leq \vec{\widetilde{\lambda}}\cdot \vec{\widetilde{\lambda}}$, which further implies the following inequality:
\begin{equation}
\vec{\lambda}\cdot\vec{\lambda}-\vec{\widetilde{\lambda}}\cdot\vec{\widetilde{\lambda}}\leq \vec{\lambda}\cdot\vec{\lambda}- \left(\sum_{i=1}^{\widehat{m}} \widetilde{\lambda}_i^2 +\frac{(1-\sum_{i=1}^{\widehat{m}} \widetilde{\lambda}_i)^2}{2^n-\widehat{m}}\right).
\end{equation}
This inequality can be combined with~\eqref{eq:eig-error5} and~\eqref{eq:eigenvecerr2} to obtain the bound in~\eqref{eq:verif}.

We now show that~\eqref{eq:verif} is tighter than~\eqref{eq:opmeaningCErr}. Specifically, we prove that the negative term in the right-hand side of~\eqref{eq:verif} is larger than the one in~\eqref{eq:opmeaningCErr}. Consider the following chain of inequalities:
\begin{align}
    \left(\sum_{i=1}^{\widehat{m}} \widetilde{\lambda}_i^2 +\frac{(1-\sum_{i=1}^{\widehat{m}} \widetilde{\lambda}_i)^2}{2^n-\widehat{m}}\right)&\geq  \sum_{i=1}^{m} \widetilde{\lambda}_i^2\nonumber\\
    &\geq \frac{(E_{m+1} - C)^2}{\sum_{i=1}^{m} (E_{m+1} - E_i)^2}\,,\nonumber
\end{align}
where we used $\widehat{m}>m$, and where the last inequality follows from~\eqref{eq:opmeaningC2}.

\subsection{Gradient of the cost function}\label{sec:grad}

Here we show that the partial derivative of \eqref{eq:costth} with respect to an angle $\theta_\nu$  is given by
\begin{align}\label{eq:gradient}
    \frac{\partial C(t,\thv)}{\partial \theta_\nu}=\frac{1}{2}\Big(&\Tr\left[H(t)V(\thv_+)\rho V\ad(\thv_+)\right]\nonumber\\
    &-\Tr\left[H(t)V(\thv_-)\rho V\ad(\thv_-)\right]\Big)\,.
\end{align}
Writing $\thv=(\theta_1,\ldots,\th_\nu,\ldots)$, then $\thv_\pm$ are simply given by $\thv_\pm=(\theta_1,\ldots,\th_\nu\pm \pi/2,\ldots)$, which shows that the gradient values are efficiently accessible by shifting the parameters in $\thv$ and measuring the expectation value  $\langle H(t)\rangle$.  

Let us consider the Layered Hardware Efficient Ansatz of Fig.~\ref{fig:ansatz}(a). Here $V(\thv)$ consists of a fixed number  $L$ of layers of $2$-qubit gates $B_\mu(\thv_\mu)$ acting on alternating pairs of neighboring qubits. Moreover, $B_\mu(\thv_\mu)$ can always be expressed  as a product of $\eta_\mu$ gates from a given alphabet $\mathcal{A}=\{U_k(\theta_k)\}$ as
\begin{equation}\label{eq:block}
    B_\mu(\thv_\mu)=U_{\eta_\mu}(\theta_\mu^{\eta_\mu})\ldots U_{\nu}(\theta_\mu^{\nu})\ldots U_{1}(\theta_\mu^{\eta_1})\,.
\end{equation}
Here $\theta_\mu^{\eta_\mu}$ are continous parameters, and we can always write without loss of generality $U_k(\theta)=R_k(\theta)T_k$, where $R_k(\theta)=e^{i \theta \sigma_k/2}$ is a single qubit rotation and $T_k$ is an unparametrized gate.

We can then compute $\partial_\nu B_\mu(\thv_\mu)\equiv \partial B_\mu(\thv_\mu)/\partial \theta_\mu^\nu$ as 
\begin{align}\label{eq:partialbeta}
    \partial_\nu B_\mu(\thv_\mu)=\frac{i}{2}  U_{\eta_\mu}(\theta_\mu^{\eta_\mu})\ldots \sigma_\nu U_{\nu}(\theta_\mu^{\nu})\ldots U_{1}(\theta_\mu^{\eta_1})\,.
\end{align}
Then, without loss of generality let us write $V(\thv)=V_L(\thv_L)B_\mu(\thv_\mu)V_R(\thv_R)$, where $V_L(\thv_L)$, and $V_R(\thv_R)$ contain  all gates in $V(\thv)$ except for $B_\mu(\thv_\mu)$. By noting that $\partial_\nu V(\thv)=V_L(\thv_L)\partial_\nu B_\mu(\thv_\mu)V_R(\thv_R)$, 
we have
\begin{align}
    \partial_\nu C=&\Tr\left[H V_L\partial_\nu B_\mu V_R\rho V_R\ad B_\mu\ad V_L\ad \right] \nonumber\\
    &+ \Tr\left[H V_L B_\mu V_R \rho V_R\ad\partial_\nu B_\mu\ad V_L\ad\right]\,,\nonumber
\end{align}
where we omitted the paramater dependence for simplicity. Then, from Eq.~\eqref{eq:partialbeta} and using the following identity (which is valid for any matrix $A$)
\begin{equation}
    i [\sigma_\nu A]=R_\nu(-\frac{\pi}{2}) A R_\nu\ad(-\frac{\pi}{2})- R_\nu(\frac{\pi}{2}) A R_\nu\ad(\frac{\pi}{2})\,,
\end{equation}
where $R_k(\theta)=e^{i \theta \sigma_k/2}$, we obtain
\begin{align}\label{eq:gradientM}
    \frac{\partial C(t,\thv)}{\partial \theta_\nu}=\frac{1}{2}\Big(&\Tr\left[H(t)V(\thv_+)\rho V\ad(\thv_+)\right]\nonumber\\
    &-\Tr\left[H(t)V(\thv_-)\rho V\ad(\thv_-)\right]\Big)\,.
\end{align}

\subsection{Algorithm for the adaptive cost function}\label{sec:alg}

Algorithm~\ref{alg1} shows a simple adaptive strategy that illustrates how one can update $H(t)$. Specifically, we consider the case when $f(t)$ is a stepwise function. In addition, we define the VQSE optimization loop termination condition in terms of the maximum number of iterations allowed $N_{\text{max}}$. We also define an updating parameter $s$ (with $N_{\text{max}}/s$ being an integer) such that we update $H_G(t)$ every $s$ steps. Finally, here we use the term optimizer, denoted as $\ensuremath{\mathsf{opt}}$, as a function that takes as inputs a set of parameters $\thv$ and a cost function $C(t,\thv)$ (or the gradient of the cost for gradient-based optimization) and returns an updated set of parameters that attempts to solve the minimization problem of~\eqref{eq:thetopt}.

\begin{figure}[t]
\begin{algorithm}[H]
\begin{itemize}[label={},leftmargin=\algorithmicindent]
\item \textbf{Input}: state $\rho$; trainable unitary $V(\thv)$; integer $m$; timestep $\delta t=1/N_{\text{max}}$; adapting stepsize $t_s=1/s$;  local time-independent Hamiltonian  $H_L$; a set of constant parameters $\{q_i\}_{i=1}^m$; classical optimizer $\ensuremath{\mathsf{opt}}$.
\item   \textbf{Output}: parameters $\thvo$ which minimize the cost  function, i.e.,  $\thvo=\arg \min_{\thv} C (\thv)$.
\item  \textbf{Init}: randomly choose a set of initial parameters $\thv$;  $H(t)\gets H_L$;  $t \gets \delta t$
\end{itemize}
\caption{Adaptive cost function with stepwise schedule $f(t)$}\label{alg1}
\begin{algorithmic}[1]
\While{$t\leq 1$}
\If {$t$ if divisible by $t_s$} 
\State {measure $V(\thv)\rho V\ad(\thv)$ in the standard basis.}
\NoNumber{define the sets  $\LC$ and  $\ZC$}
\State {$H_G(t)\gets \id-\sum_{i=1}^{m} q_i\ketbra{\zv_i}{\zv_i}$} with $z_i\in \ZC$
\State {$H(t)\gets (1-t)H_L+t H_G(t)$}
\EndIf
\State {run $\ensuremath{\mathsf{opt}}$ with  $C$ and $\thv$ as input, and $\thv_{\text{min}}$ as output}
\State {$\thv\gets \thv_{\text{min}}$}
\State {$t\gets t +\delta t$}

\If {$t =1$}
\State {$\thvo\gets \thv$}
\EndIf

\EndWhile

\hspace*{-\algorithmicindent} \textbf{Return}: {$\thvo$}

\end{algorithmic}
\end{algorithm}
\end{figure}

\section{Data availability}
Data generated and analyzed during current study are available from the corresponding author upon reasonable request

\section{Acknowledgements}
We thank Lukasz Cincio for helpful conversations. All authors acknowledge support from LANL's Laboratory Directed Research and Development (LDRD) program. MC was also supported by the Center for Nonlinear Studies at LANL. PJC also acknowledges support from the LANL ASC Beyond Moore's Law project. This work was also supported by the U.S. Department of Energy (DOE), Office of Science, Office of Advanced Scientific Computing Research, under the Quantum Computing Application Teams program, and by the U.S. DOE, Office of Science, Basic Energy Sciences, Materials Sciences and Engineering Division, Condensed Matter Theory Program.

\section{Competing Interests}
The authors declare no competing interests.

\section{Author Contribution}
The project was conceived by PJC. The manuscript was written by MC, KS, AA,  PJC. The theoretical results were derived by MC, KS, and PJC.   MC, and KS performed the numerical simulations in Fig. 4, 5, and 7. AA developed the noisy simulator and performed the numerical simulations for error mitigation of Fig. 6.  MC and KS contributed equally to this work and are considered as co-first authors.


\bibliography{ref.bib}

\end{document}